\documentclass[letterpaper,10pt,twocolumn]{article}

 \title{BMS: Secure Decentralized Reconfiguration for Blockchain and BFT Systems}
 \author{
    Selma Steinhoff \\ ETH Zurich\thanks{Work done in IBM Research Europe - Zurich} \and
    Chrysoula Stathakopoulou \\IBM Research Europe - Zurich \\ ETH Zurich \and
    Matej Pavlovic \\ IBM Research Europe - Zurich \and
    Marko Vukoli\'c \\Protocol Labs\footnotemark[1]
    \date{}
 }

\usepackage{graphicx}
\usepackage{amsfonts}
\usepackage[outdir=./]{epstopdf}

\usepackage[utf8]{inputenc}
\usepackage{amsmath}
\usepackage{subcaption}
\usepackage{algorithm}
\usepackage[noend]{algpseudocode}
\usepackage{epigraph}
\usepackage{etoolbox}
\usepackage{graphicx}
\usepackage{float}
\usepackage{afterpage}
\usepackage{pdfpages}
\usepackage{array}
\usepackage{url}
\makeatletter
\g@addto@macro{\UrlBreaks}{\UrlOrds}
\makeatother

\makeatletter
\patchcmd{\epigraph}{\@epitext{#1}}{\itshape\@epitext{#1}}{}{}
\makeatother

\makeatletter
\def\BState{\State\hskip-\ALG@thistlm}
\makeatother

\newcolumntype{P}[1]{>{\centering\arraybackslash}p{#1}}

\usepackage{filecontents}
\usepackage{graphicx}
\usepackage{hhline}
\usepackage{xspace}
\usepackage{ulem}
\usepackage{tikz}
\usepackage{amsmath}
\usepackage{enumerate}
\usepackage{enumitem}

\usepackage{algorithmicx}
\usepackage{algorithm}
\usepackage{algpseudocode}
\usepackage{varwidth}
\usepackage{gensymb}
\usepackage{makecell}
\usepackage{tabularx, booktabs}
\usepackage{array}
\usepackage{times}

\newcommand{\omitit}[1]{}

\usepackage{mathtools}

\algnewcommand{\LineComment}[1]{\Statex \hskip\ALG@thistlm \(\triangleright\) #1}

\newcommand{\leavereq}{\textsc{LEAVE-REQUEST}\xspace}
\newcommand{\registertx}{\textsc{REGISTER}\xspace}
\newcommand{\registerannounce}{\textsc{REGISTER-ANNOUNCE}\xspace}
\newcommand{\registerconfirm}{\textsc{REGISTER-CONFIRM}\xspace}
\newcommand{\joinreq}{\textsc{JOIN-REQUEST}\xspace}

\newcommand{\vote}{\textsc{VOTE}\xspace}
\newcommand{\configreq}{\textsc{CONFIG-REQUEST}\xspace}
\newcommand{\configres}{\textsc{CONFIG-RESPONSE}\xspace}

\newcommand{\event}[3]{
\ifthenelse
{\equal{#3}{}}
{\ap{#1.\textrm{#2}}}
{\ap{#1.\textrm{#2} \mid #3}}
}

\algnewcommand\Instance[2]{\State #1, \textbf{instance} #2}
\algnewcommand\InstanceSystem[3]{\State #1, \textbf{instance} #2, \textbf{system} #3}
\algnewcommand\Trigger[3]{\State \textbf{trigger} $\event{#1}{#2}{#3}$}
\algnewcommand\Schedule[1]{\State \textbf{schedule} $#1$}
\algnewcommand\Cancel[1]{\State \textbf{cancel} $#1$}
\algnewcommand\Broadcast[1]{\State \textbf{broadcast} $#1$}
\algnewcommand\Reply[1]{\State \textbf{reply} $#1$}
\algnewcommand\Import[1]{\State \textbf{import} $#1$}
\algnewcommand\Not{\textbf{ not }}
\algnewcommand\AndT{\textbf{ and }}
\algnewcommand\OrT{\textbf{ or }}
\algnewcommand\In{\textbf{ in }}

\algsetblockdefx[Implements]{Implements}{EndImplements}{}{}{\textbf{Implements:}}{}
\algsetblockdefx[Uses]{Uses}{EndUses}{}{}{\textbf{Uses:}}{}
\algsetblockdefx[Parameters]{Parameters}{EndParameters}{}{}{\textbf{Parameters:}}{}
\algsetblockdefx[Init]{Init}{EndInit}{}{}{\textbf{Init:}}{}
\algsetblockdefx[Event]{Event}{EndEvent}{}{}{\textbf{Event:}}{}

\algsetblockdefx[Upon]{Upon}{EndUpon}{}{}[1]{\textbf{upon} $#1$ \textbf{do}}{}
\algsetblockdefx[UponExists]{UponExists}{EndUponExists}{}{}[2]{\textbf{upon exists} $#1$ \textbf{such that} $#2$ \textbf{do}}{}
\algsetblockdefx[UponCondition]{UponCondition}{EndUponCondition}{}{}[1]{\textbf{upon} $#1$ \textbf{do}}{}
\algsetblockdefx[UponReceiving]{UponReceiving}{EndUponReceiving}{}{}[1]{\textbf{upon receiving} $#1$ \textbf{do}}{}
\algsetblockdefx[UponCommitting]{UponCommitting}{EndUponCommitting}{}{}[1]{\textbf{upon committing} $#1$ \textbf{do}}{}

\algsetblockdefx[Procedure]{Procedure}{EndProcedure}{}{}[2]{\textbf{procedure} \emph{#1}($#2$) \textbf{is}}{}
\algsetblockdefx[Function]{Function}{EndFunction}{}{}[2]{\textbf{function} \emph{#1}($#2$) \textbf{:}}{}
\algsetblockdefx[FunctionNoArgs]{FunctionNoArgs}{EndFunctionNoArgs}{}{}[1]{\textbf{function} \emph{#1}() \textbf{:}}{}

\algdef{SE}[DoUntil]{DoUntil}{EndDoUntil}{\textbf{do}}[1]{\textbf{until} $#1$}
\algsetblockdefx[Struct]{Struct}{EndStruct}{}{}[1]{\textbf{Struct} $#1$ \textbf{contains}}{}
\algsetblockdefx[IfExists]{IfExists}{EndIfExists}{}{}[2]{\textbf{if exists} $#1$ \textbf{such that} $#2$ \textbf{then}}{\textbf{end if}}
\algsetblockdefx[While]{While}{EndWhile}{}{}[1]{\textbf{while} $#1$ \textbf{do}}{\textbf{end while}}
\algsetblockdefx[NTimes]{NTimes}{EndNTimes}{}{}[1]{\textbf{for} $#1$ \textbf{times do}}{\textbf{end for}}
\algsetblockdefx[For]{For}{EndFor}{}{}[3]{\textbf{for} $#1 \in #2..#3$ \textbf{do}}{\textbf{end for}}
\algsetblockdefx[ForAll]{ForAll}{EndForAll}{}{}[2]{\textbf{for all} $#1 \in #2$ \textbf{do}}{\textbf{end for}}

\begin{document}

\maketitle              

\begin{abstract}
Reconfiguration of long-lived blockchain and Byzantine fault-tolerant (BFT) systems poses fundamental security challenges.
In case of state-of-the-art Proof-of-Stake (PoS) blockchains,
stake reconfiguration enables so-called \emph{long-range attacks},
which can lead to forks.
Similarly, permissioned blockchain systems, typically based on BFT,
reconfigure internally, which makes them susceptible to a similar \emph{``I still work here''} \cite{aguilera2010reconfiguring} attack.

In this work, we propose BMS (Blockchain/BFT Membership Service)
offering a secure and dynamic reconfiguration service for BFT and blockchain systems,
preventing long-range and similar attacks.
In particular:
(1) we propose a root BMS for permissioned blockchains, implemented as an Ethereum smart contract and evaluate it reconfiguring the recently proposed Mir-BFT protocol \cite{stathakopoulou2019mir},
(2) we discuss how our BMS extends to PoS  blockchains and how it can reduce PoS  stake unbonding time from weeks/months to the order of minutes,
and (3) we discuss possible extensions of BMS to hierarchical deployments as well as to multiple root BMSs.
\end{abstract}

\section{Introduction}

Blockchain systems are conceived as long-lived systems, implemented typically by a set of \emph{nodes} (called e.g., miners, validators) who participate in consensus on system state.
More lightweight \emph{client} nodes (e.g., wallets) occasionally access blockchain nodes, e.g., when a client wishes to submit an operation to the blockchain.
In a long-lived system, this leads to a fundamental security challenge: \emph{how can a client, who might have been disconnected for a while, trust the state of the system}?

\begin{figure*}[ht!]
        \centering
	\begin{subfigure}{0.8\textwidth}
		\centering
		\includegraphics[width=\textwidth]{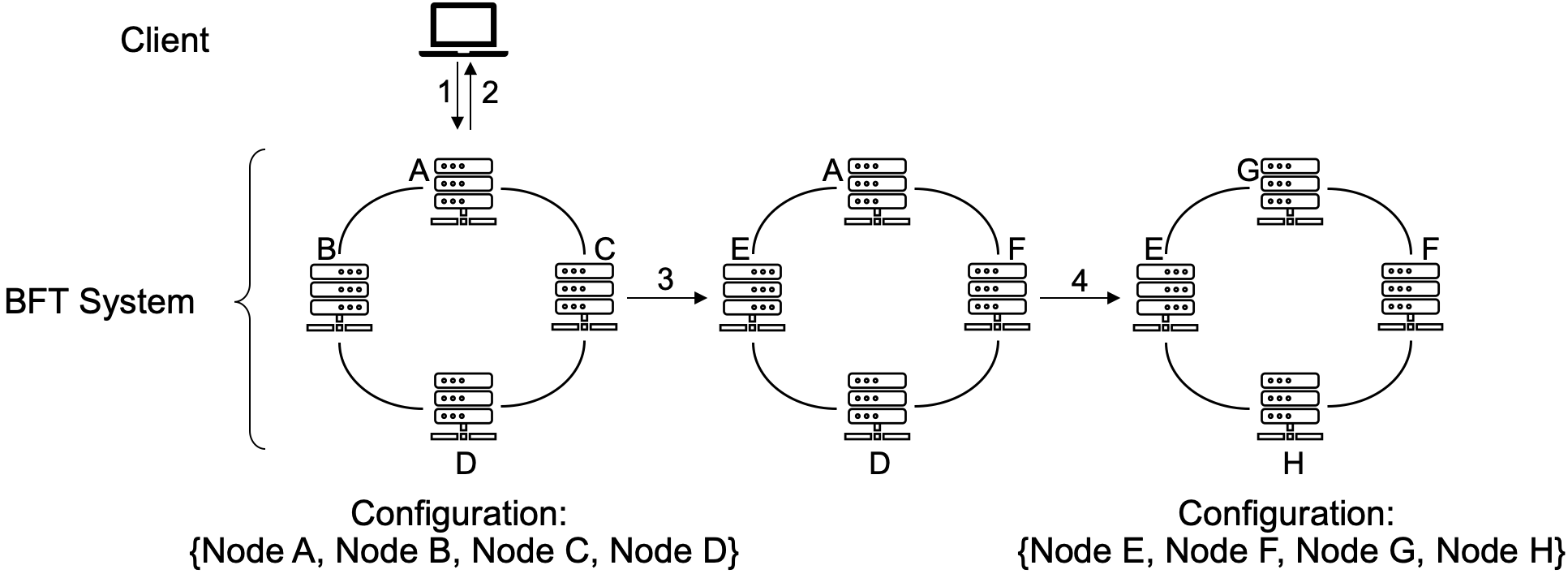}
		\caption{Client communicates with initial configuration $C_0$ which then transitions, perhaps gradually, to configuration $C_2$ (steps 1-4).}
		\label{fig:BMS-reconfig}
	\end{subfigure}\\
	\begin{subfigure}{0.8\textwidth}
		\centering
		\includegraphics[width=\textwidth]{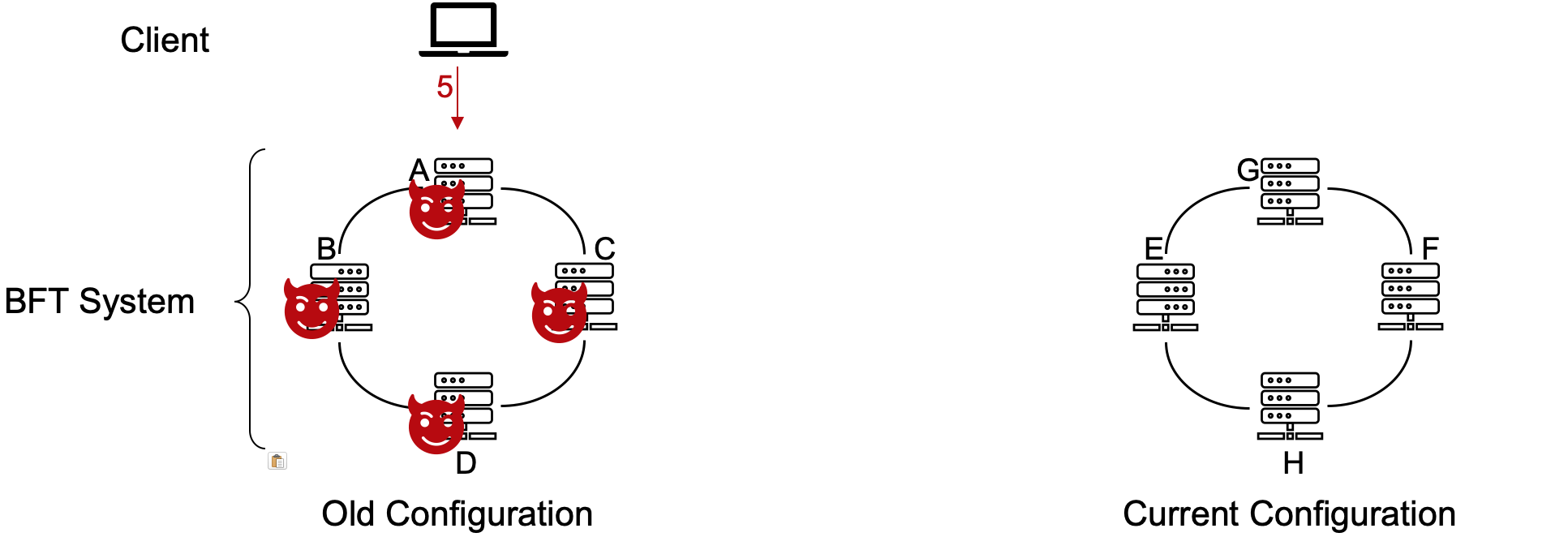}
		\caption{When client reconnects, the initial configuration $C_0$ may be entirely corrupted by the adversary (step 5).}
		\label{fig:BMS-attack}
	\end{subfigure}
	\caption{Illustration of the long-range (``I still work here'') family of attacks.}
	\label{fig:long-fork}
\end{figure*}

To visualize the issue, consider the illustration in Figure~\ref{fig:long-fork}.
In this example, the initial membership (configuration $C_0$) comprising nodes A, B, C and D has reconfigured entirely to a new configuration $C_2$
which includes no nodes from $C_0$ (Fig.~\ref{fig:BMS-reconfig}).
A client that reconnects only occasionally to the blockchain system may have not been updated about changes in the system membership,
and reconnect to the initial configuration $C_0$ (Fig.~\ref{fig:BMS-attack}).
However, the initial configuration might be entirely corrupted by the adversary
as realistic blockchain systems limit assumptions on the power of the adversary only with respect to their current configuration.

This issue affects \emph{all} blockchain systems, whether they are based on Proof-of-Work (PoW), on Proof-of-Stake (PoS), or are permissioned.
In PoS blockchains, the adversary can at this point mount a long-range attack \cite{DeirmentzoglouP19,AzouviDN19}
fabricating a blockchain history which the client cannot distinguish from the ``real'' one.
A similar attack is known in permissioned blockchains and Byzantine fault-tolerant (BFT) systems as the ``I still work here'' attack \cite{aguilera2010reconfiguring}.

While the issue depicted in Figure~\ref{fig:long-fork} affects PoW blockchains as well, these can rely --- unlike PoS and permissioned blockchains --- on the longest chain selection rule \cite{Bitcoin}.
This, together with ever growing miner hashing power in established networks such as Bitcoin and Ethereum,
enables nodes of the current configuration to (eventually) convince the client of their legitimacy.
This powerful security feature of PoW blockchains presents their critical advantage over PoS and permissioned blockchain systems.

In PoS systems, long-range attacks are typically partially addressed by making validators bond their stakes for a very long time (stake unbonding or ``thawing'' time, which is often several weeks \cite{CosmosThawing} or more) with slashing mechanisms to discourage adversarial behavior. Other approaches require clients to periodically synchronize with changes in the membership set \cite{AzouviDN19}. As for permissioned blockchains, they typically reconfigure internally and are as such entirely vulnerable to this type of attacks.

In this paper, we take a different approach and propose a general decentralized solution for secure and dynamic reconfiguration for blockchain and BFT distributed systems. This paper presents the following key contributions:

\textit{1.} We propose BMS, a Blockchain / BFT Membership Service for permissioned blockchains and BFT systems which prevents the ``I still work here'' attack by leveraging a (root) BMS implemented as a smart-contract on a public permissionless PoW blockchain. Members of the BFT distributed system push membership updates to the BMS; the new membership is accepted and published by the BMS once a qualified number of nodes of the reconfigured system approve (vote for) the change. Our BMS allows the new permissioned blockchain nodes to pay fees for joining and leaving the system. These fees are distributed among voting nodes to cover the cost of pushing membership updates to the BMS and create incentives for protocol participation.

\textit{2.} We implement a BMS prototype as a smart-contract on the Ropsten Ethereum testnet \cite{ropsten} and use it with the Mir-BFT protocol \cite{stathakopoulou2019mir}.
We evaluate the latency and cost of our reconfiguration scheme for increasing system sizes up to 100 nodes.
We also evaluate the impact of different reconfiguration policies.

\textit{3.} Finally, we discuss how to extend our BMS from permissioned to PoS blockchains to prevent  the long-range attack, where BMS can considerably reduce the stake unbonding (``thawing'') time. We also discuss possible extensions of BMS to hierarchical deployments (akin to Internet DNS) as well as to multiple root BMSs.

In the rest of this paper, we present the system model (Section \ref{sec:model}),
give an overview of BMS (Section \ref{sec:overview}),
present the details of our protocols (Section \ref{sec:alg-details}),
evaluate our BMS prototype implementation (Section \ref{sec:evaluation}),
discuss PoS and other extensions of BMS (Section \ref{sec:extensions}),
and compare BMS to related work (Section \ref{sec:related}).

\section{System Model}
\label{sec:model}

We consider a message-passing system and an

eventually synchronous network \cite{Dwork:1988:CPP:42282.42283}
in which messages sent between any processes before an (unknown) Global Stabilization Time (GST)
can be delayed arbitrarily or dropped.
However, messages sent between correct processes after GST are guaranteed to have bounded delay.

We define a \emph{reconfigurable distributed Byzantine fault-tolerant} system
(referred to, simply, as the BFT system)
as any distributed system composed of multiple \emph{member nodes}
that execute a distributed application.
We assume that a certain fraction (see below) of the nodes can be Byzantine faulty \cite{BGP},
but even faulty processes cannot compromise standard cryptographic primitives such as secure hash functions.
We model the application as a replicated state machine \cite{Schneider90} across these nodes.
The application state can be updated through requests that \emph{clients} submit to the system
(by sending request messages to the member nodes).
Each node maintains a copy of a totally ordered log of such requests
and applies them to its copy of the application state in log order.

The BFT system being reconfigurable means that the set of its member nodes,
that we call the \emph{configuration},
may dynamically change.
The system starts in an initial configuration $C_0$ that all nodes and all clients are aware of.
During operation, the system passes through a sequence of configurations $C_0, C_1, C_2, \dots$ and so on,
as nodes join and leave the system.
We assume that the BFT system reconfigures autonomously
by maintaining its configuration alongside its application state.
A transition from $C_k$ to $C_{k+1}$ happens at
agreed-upon positions in the request log that we call \emph{checkpoints}.
All nodes use configurations $C_k$ and $C_{k+1}$
for ordering requests that occur in the log respectively before and after the checkpoint.

In a configuration $C$, we assume that up to $f(C)<\frac{|C|}{3}$ nodes can be Byzantine faulty,
if $C$ is the last configuration or if no newer configuration has been published by the BMS,
or if a newer configuration has been published only recently.
``Recently'' means before time $P$ (publishing time) has elapsed since publishing the new configuration by BMS.

Here we need to slightly depart from the standard definition of a correct node as one that \emph{never} misbehaves.
Instead, we regard node $q$ as correct even if $q$ misbehaves, as long as this happens $P$ units of time
after the BMS publishes a configuration $q$ is not part of any more.
We call a node that is not correct faulty.

We further assume that a correct client's view of the BMS state is not older than $P$
whenever the client interacts with the BFT system.
Note that, without BMS, the client might be required to maintain
a sufficiently up-to-date view of the BFT system's configuration
at all times (not just while interacting with the BFT system), as is the case in other systems \cite{AzouviDN19,kwon2016cosmos}.
Weakening this requirement is an important part of this paper's contribution.

In order to interact with the BFT system, clients must be aware of the BFT system's current configuration.
We only assume, however, that clients are aware of the initial configuration $C_0$,
and it is the protocol's responsibility to keep clients' view of the configuration up to date.

In addition to the BFT system (that implements a replicated state machine),
we assume the existence of another external public state machine replication (SMR) service that is \emph{reliably discoverable}.
This means that a client can securely use the service
without any prior information about the system other than what is public knowledge.
Ethereum is an example of such an SMR system, since,
as a PoW system, it does not suffer from long-range attacks.
We use this external SMR system to implement BMS.

\section{Overview of BMS}
\label{sec:overview}

On a high level, the system is composed of two parts: (1) the BFT system itself, and (2)
the reconfiguration service (BMS), tracking the membership of the former.

A client accessing the BFT system is required to know its configuration (network addresses, keys, etc.).
A crucial property of BMS
is that it is accessible without any prior knowledge about the identities of the nodes running it.
Public PoW-based blockchain systems like Ethereum satisfy this property.
If the reconfiguration service has the form of an Ethereum smart contract (as in our implementation),
the identity of this smart contract can be considered the identity of the whole system,
as it is the only information needed to bootstrap interaction with our system.

In a nutshell, a client that needs to connect to the BFT system
first reliably obtains an up to date configuration of the BFT system from the BMS.
The client then submits its requests to the BFT system.

In the following, we describe the state machines executed by the respective distributed systems and their interaction.
We first describe a simplified version of the system (Sec. \ref{sec:overview-service} and \ref{sec:overview-bft-system}).
Then, we refine it to account for a potential performance bottleneck imposed by the reconfiguration service
and real-world cost of using the reconfiguration service---%
in our case of Ethereum, the Gas cost of transactions (Sec. \ref{sec:overview-cost}).

\subsection{Reconfiguration Service}
\label{sec:overview-service}

The reconfiguration service is a state machine executed by a system
whose identity is public knowledge and is hard to forge.
A public PoW-based permissionless blockchain platform like Ethereum is an example of such a system.

BMS does not necessarily need to be implemented on Ethereum, it is only important that BMS
is able to execute the logic of a state machine driven by client requests,
often called transactions.  For clarity, we use the term ``request'' for requests submitted to the BFT system, while we refer to operations invoked on BMS as ``transactions''.

The state of BMS consists of the configuration of the BFT system,
initialized to its initial configuration.
This configuration can only be updated if sufficiently many members of it
submit a specific transaction (that we call a \emph{vote}) containing the same new configuration.
For configuration $C_i$ stored at the BMS, ``sufficiently many'' corresponds to $f(C_i) + 1$,
guaranteeing that at least one of the voting nodes is correct.
Only members of $C_i$ can act as voting clients to BMS.
The service ignores all votes not signed by members of $C_i$.

As soon as the service receives enough votes for a new configuration $C_j$,
it replaces the old configuration $C_i$ with $C_j$ and waits for new votes
(this time from nodes in $C_j$)
before it updates the configuration again. Any process can, however,
query the reconfiguration service for the current configuration.
This is used by new nodes joining the BFT system and clients (of the BFT system)
that want to submit requests to the BFT system.

\subsection{Reconfigurable BFT System}
\label{sec:overview-bft-system}

The BFT system implements a replicated state machine.
The state of this state machine also includes
(in addition to the state of the executed application)
the BFT system's configuration, i.e., the identities (network addresses, public keys)
of the nodes executing it.
The state of the state machine is updated by processing an ordered log of requests submitted by clients.
In order to join the system,
a node first acts as a client to the BFT system
and submits a special reconfiguration request.

The nodes in the current membership treat this request as they would any other client request,
until two conditions are satisfied: (1) the reconfiguration request is inserted in the request log
	and all preceding requests have been processed, and (2) the BMS is up to date,
	i.e., the current configuration is reflected in the state of the reconfiguration service.

Once both conditions hold, the BFT system reconfigures.
This comprises each node updating the current configuration in its local state to a new one
and sending a response to the joining node.
The response contains all information the joining node needs to start executing the BFT protocol:
the application state, the new configuration, and all protocol-specific data.
We use an analogous procedure for nodes leaving the system.

After the BFT system reconfigures,
each node in the \emph{old} configuration (the one still saved by the reconfiguration service)
acts as a client to the reconfiguration service and submits its vote
(as a transaction to BMS) with the new configuration (see also Sec.~\ref{sec:overview-service}).

\subsection{Batching and Incentives}
\label{sec:overview-cost}

Our BMS is intended to be hosted on a public PoW blockchain platform like Ethereum.
However, such platforms usually imply high transaction latencies (minutes to hours)
and a real-world cost in form of transaction fees (expressed as ``Gas'' in Ethereum).
This has major implications on the reconfiguration protocol as described so far: (1) churn (i.e. the rate at which nodes can join and leave the BFT system)
is limited to the inverse of the transaction latency, and (2) a join or leave event of a node incurs additional financial cost for the other members of the configuration
that send their vote transactions to the reconfiguration service and have to assume the corresponding transaction fees.

In the following, we present two modifications to the algorithm described above that
(1) enable higher churn rates in exchange for stronger failure assumptions (\emph{batching}) and
(2) make each node bare the transaction fees resulting from its joining and leaving the system (\emph{incentives}).

\subsubsection{Batching Updates to BMS}
\label{sec:batching}
In order to allow higher churn,
the BFT system need not notify the reconfiguration service about every single configuration change.
Instead, the BFT system may reconfigure locally, without pushing all configurations to BMS,
and only when the configuration of the BFT system becomes ``too different'' from the one stored by BMS,
or only when a certain amount of (logical) time passes%
\footnote{Logical time in a BFT system can be expressed in the number of requests processed,
while enforcing a minimal request rate by injecting special ``noop'' requests as necessary.},
the BFT system updates the BMS state.

As we do not expect nodes which left the system to participate in voting,
there always needs to be a sufficient overlap between the last published configuration and the local configuration of the BFT system.
Otherwise the BFT system looses the ability to publish new configurations to the BMS.
Since $f(C_i) + 1$ correct nodes of the \emph{last published} configuration $C_i$
are necessary for updating the BMS state, we require for any subsequent unpublished local configuration $C_j$ of the BFT system
that $|C_i \cap C_j| \geq f(C_i) + f(C_j) + 1$.
This way, even if all $f(C_j)$ nodes in $C_j$ misbehave, $C_j$ will still contain sufficiently many $(f(C_i) + 1)$ correct nodes
to update the BMS state.

In practice, this can easily be implemented
if the state machine executed by the BFT system keeps track of the number of join / leave events
that occurred since it pushed the last configuration to BMS.
As long as this number is below a configured threshold $t$ (i.e., not ``too different''),
the BFT system does not enforce condition (2) of Section \ref{sec:overview-bft-system}
and it does not notify the reconfiguration service.
The simplified case described so far corresponds to $t = 1$,
where even a single join/leave event is always announced to BMS.

The threshold $t$ is limited by the required overlap between the last published configuration $C_i$
and the local configuration $C_j$ of the BFT system, as discussed above.
As configuration $C_i$ is, by assumption, only guaranteed to contain $|C_i| - f(C_i)$ correct nodes,
no more than $|C_i| - f(C_i) - (f(C_i) + 1) = |C_i| - (2f(C_i) + 1)$ correct nodes can leave the system
before updating the published configuration becomes impossible.
This, together with the consequences of always maintaining more than two thirds of honest nodes in every individual configuration,
leads to a bound $t \leq \lfloor 3/2 \cdot f(C_i) \rfloor + 1 + ((|C_i|-1) \mathtt{\ mod\ } 3)$%
\footnote{The term $((|C_i|-1) \mathtt{\ mod\ } 3)$ accounts for configurations $C_i$ with $|C_i| > 3 \cdot f(C_i) + 1$.
  Those configurations contain one or two ``additional'' nodes that contribute to a greater system size $|C_i|$,
but are not sufficient to enable tolerating an additional node failure.}.
In particular, for configurations with an optimal size with respect to fault tolerance, where $|C_i| = 3 \cdot f(C_i) + 1$,
this translates to $t \leq \lceil |C_i|/2 \rceil$.

Increasing $t$ also results in an according increase of the period during which a departing node is assumed to not misbehave.
This is implied by the assumption that less than one third of nodes misbehave in every configuration
\emph{until a newer configuration is published by the BMS} (plus the publishing time $P$).
In other words, we assume sequences of $t$ consecutive configurations
such that in each of those configurations no more than one third of nodes misbehave
until the last of those configurations is published by the BMS.

\subsubsection{Transaction Costs and Incentives.}
To make each node pay for the cost pertaining to its reconfiguration operations,
we add a \emph{registration} mechanism to BMS, where we
require each node joining the system
to pay a registration fee.
The BMS then uses this fee
to reimburse members of the BFT system for their voting transactions.

The form of the registration fee is platform-dependent
and is expressed in the same way as the cost of executing transactions on that platform.
In the case of Ethereum, for example, the cost of executing a transaction is measured in Gas
(or Ether, that can be converted to Gas),
and so is the registration fee.
To register, the joining node transfers the registration fee
to the smart contract implementing BMS.

Only after the joining node is registered with the BMS,
it submits its reconfiguration request to the BFT system.
The BFT system only handles the reconfiguration request
after verifying that the joining node has registered with BMS and paid the registration fee.
When it comes to voting on the new configuration that includes the joining node,
BMS reimburses the voting members using the joining node's registration fee.
Concretely on Ethereum, the voting transaction performed by a member of the BFT system
also induces a transfer of an appropriate part of the joining node's registration fee
from the smart contract to the voting member's Ethereum account.

The registration fee is not fully redistributed to voting members during the process of joining.
The reconfiguration service retains part of the registration fee
to analogously cover the costs associated with a node \emph{leaving} the BFT system.

\section{Algorithm Details}
\label{sec:alg-details}

\begin{figure*}[h]
	\centering
	\includegraphics[width=\textwidth]{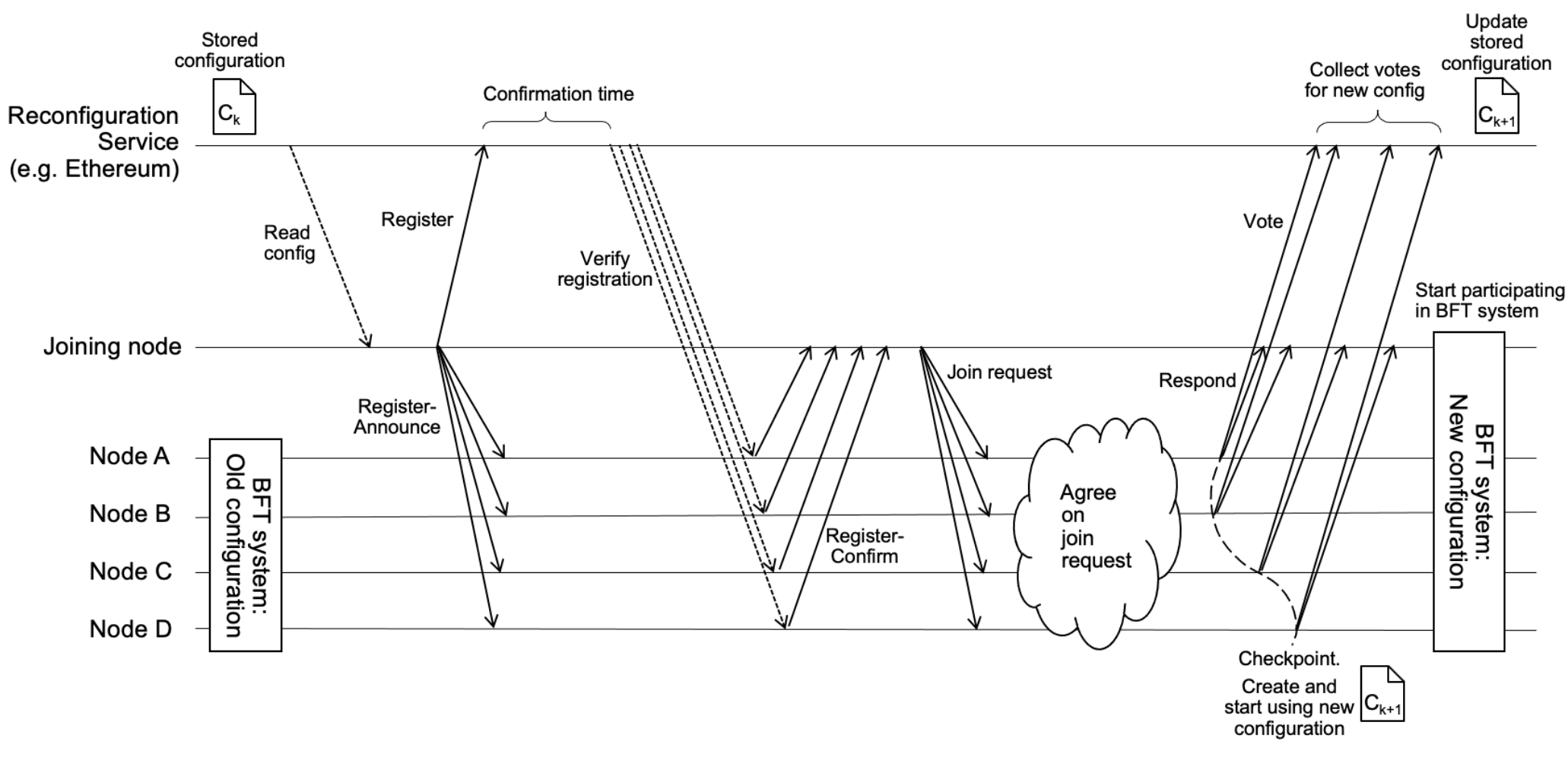}
	\caption{Execution of our protocol for adding a new node to the configuration.
	The BFT system's configuration $C_k$ transitions to configuration $C_{k+1}$, augmented by the joining node.
	The transfers of funds within the reconfiguration service happen on processing the ``Register'' transaction
	(from the joining node's account to the BMS smart contract)	and on updating the stored configuration
	(from the smart contract to the accounts of nodes which voted for $C_{k+1}$).}
	\label{fig:protocol-diagram}
\end{figure*}

This section presents the protocols of our system in detail.
Algorithm~\ref{algorithm:ReconService} describes the logic of the state machine implementing the reconfiguration service (BMS) itself.
Algorithm~\ref{algorithm:ReconProt} defines the extension to the BFT system for interaction with joining/leaving nodes and the reconfiguration service.
It is executed by each member of the BFT system in addition to the protocols implementing the BFT system itself.

The interaction between MBS and the BFT system is depicted in Figure \ref{fig:protocol-diagram}
using an example execution of a new node joining the BFT system.

We assume that all nodes have access to the state of the reconfiguration service.
In particular, for an Ethereum-based implementation of BMS, each node is assumed to run its own Ethereum client (or use a proxy).
We also assume that all messages and requests are authenticated using an appropriate authentication mechanism, e.g., public-key cryptography.

\begin{algorithm*}
	\scriptsize
	\begin{algorithmic}[1]

		\Parameters
		\State $cost$ \Comment{Joining fee}
		\EndParameters

		\Struct{Configuration}
		\State $number$   \Comment{Configuration number}
		\State $members$  \Comment{List of member identifiers}
		\State $v$        \Comment{Threshold of votes for accepting the next configuration}
		\EndStruct

		\Init
		\State $C_{cur}\leftarrow C_0$ \Comment{Current configuration}
		\State $Registrations\leftarrow \emptyset$  \Comment{Set of valid registrations}
		\State $Votes\leftarrow \{\}$ \Comment{Map from configurations to sets nodes voting for them}
		\EndInit

		\State \label{ln:regStart}\textbf{upon receiving } $\langle\registertx,id,fee\rangle$
		\State                    \hskip 0.55cm \textbf{ such that} $fee \geq cost $\textbf{ do}
		\State   \label{ln:regEnd}\hskip 0.55cm \hskip \algorithmicindent  $Registrations\leftarrow Registrations\cup \{r\}$ \Comment{Store registrations}
		\State

		\State \textbf{upon receiving } $\langle\vote,C,p\rangle$\textbf{ from} $p$
		\State \label{ln:checkMember}\hskip 0.55cm \textbf{ such that} $p\in C_{cur}.members$\textbf{ do}
		\State \label{ln:updateVotes}\hskip 0.55cm \hskip \algorithmicindent  $Votes[C]\leftarrow Votes[C]\cup \{p\}$ \Comment{Store votes}
		\State

		\State \label{ln:voteCount}\textbf{upon } $\exists C : C.number > C_{cur}.number \wedge |Votes[C]| \geq C_{cur}.v$
		\State
                \State \hskip 0.55cm $reward \leftarrow 0$                                                                \Comment{Compute reward ($\triangle$ denotes symmetric set difference):}
		\State \hskip 0.55cm \textbf{for all } $r \in Registrations : r.id \in (C \triangle C_{cur})$\textbf{ do} \Comment{For each joining or leaving node}
                \State \hskip 0.55cm \hskip \algorithmicindent $reward \leftarrow reward + (r.fee / 2)$                   \Comment{Add half of registration fee to reward}
		\State
		\State \hskip 0.55cm \textbf{for all } $p \in Votes[C]$\textbf{ do}                  \Comment{Reward voters}
		\State \label{ln:sendReward}\hskip 0.55cm \hskip \algorithmicindent transfer $(reward/|Votes[C]|)$ to $p$
		\State
		\State \label{ln:updateConfig}\hskip 0.55cm $C_{cur}\leftarrow C$ \Comment{Update configuration}

	\end{algorithmic}
	\caption{BMS}
	\label{algorithm:ReconService}
\end{algorithm*}

\subsection{BMS Algorithm (Algorithm~\ref{algorithm:ReconService})}
\label{sec:ReconServiceDesign}
BMS stores the configuration $C_i$ of the BFT system and the parameter $cost$ which specifies the registration fee.
BMS is public, so anyone can read its state and learn the configuration $C_i$ of the BFT system.
A read operation from a process to the reconfiguration service has the form $\langle\configreq\rangle$
and returns a $\langle\configres, C_i\rangle$, where $C_i$ is the stored configuration.
We omit this part in the pseudocode for brevity.

BMS records registration transactions sent by joining nodes (lines \ref{ln:regStart}-\ref{ln:regEnd}).
A registration transaction from node $p$ has the form $\langle\registertx,p,a\rangle$,
where $a$ indicates the amount of cryptocurrency that is transferred to BMS with the transaction.
A \registertx transaction is only considered by BMS if $a \geq cost$.

The configuration $C_i$ stored in BMS can be updated to a new configuration $C_{j}$, $j > i$, in a process that we call ``voting''.
To update the configuration, a threshold of $v_i = f(C_i) + 1$ nodes from $C_i$ need to send a vote proposing the same new configuration $C_{j}$ to BMS.
A vote from a node $p$ has the form $\langle\vote, C_{j}, p\rangle$, where $C_{j}$ contains the node identifiers (including public keys) of all members of the new configuration.
BMS only accepts a \vote from a node $p$ if $p \in C_i$, i.e., if $p$ is member of the currently stored configuration (line~\ref{ln:checkMember}).
Only a single vote per node per configuration is possible (line~\ref{ln:updateVotes}).

Once BMS has recorded $v_i$ \vote transactions for a new configuration $C_j$ (line~\ref{ln:voteCount}),
it rewards the participating voters (line~\ref{ln:sendReward})
and updates the stored configuration $C_i$ to $C_j$ (line~\ref{ln:updateConfig}).
The joining fees collected through registrations of nodes which are now included in (or excluded from) the new configuration
are distributed evenly among the nodes that voted for the accepted configuration.

\subsection{Extensions to the BFT System}
\label{sec:ReconProt}

\subsubsection{Submitting Reconfiguration Requests}

When a node $p$ wants to join the BFT system,
it first reads the last published configuration $C_i$ from BMS and registers
as described in Section~\ref{sec:ReconServiceDesign}. Node $p$ then
obtains a \emph{proof of registration}.
To this end,  $p$ sends a $\langle\registerannounce,p\rangle$ message to all nodes $q \in C_i$ of the BFT system.
Each member $q$ of the BFT system receiving such a message waits until it observes $p$'s registration in the BMS state
and responds to $p$ with a $\langle\registerconfirm, p, \sigma_q\rangle$ message, with $\sigma_q$ being $q$'s signature.
A set of $f(C_i)+1$ such signed confirmations that $p$ obtains constitute a \emph{valid} proof of registration.

The joining node $p$ then (acting as a client of the BFT system) submits a special reconfiguration request $\langle\joinreq, pr\rangle$ to the BFT system.
The reconfiguration request contains the previously obtained proof of registration $pr$.
Its purpose is to convince members of the BFT system that $p$ indeed registered (and paid the registration fee),
even if some nodes have not yet observed $p$'s registration directly in the reconfiguration service.
Join requests with invalid $pr$ are ignored by correct nodes.

Similarly, a node $p$ that is currently member of the BFT system and wants to leave,
submits a leave request of the form $\langle\leavereq,p,\sigma_p\rangle$ (acting as a client to the BFT system).
The signature of the leaving node is necessary to prevent malicious nodes from evicting correct ones.
Members of the BFT system ignore leave requests with invalid signatures. Alternatively, if the BFT system possesses a mechanism for detecting misbehaving nodes,
a node different from $p$ can submit a request $\langle\leavereq,p,PoM(p),\rangle$, where $PoM(p)$ is a Proof of Misbehavior of node $p$.
This way, misbehaving nodes may actively be evicted from the system.

\begin{algorithm*}
	\scriptsize
	\begin{algorithmic}[1]
		\Import TotalOrderBroadcast
		\Import BMS

		\Parameters
		\State $id$ \Comment{The node id}
		\State $t$  \Comment{Tolerated difference between local configuration and current configuration in BMS}
		\State $f$  \Comment{Fault tolerance of the current configuration}
		\EndParameters

		\Struct{ProofOfRegistration}
		\State $confirmations$   \Comment{Set of signed registration confirmations}
		\EndStruct

		\Init
		\State $C_{cur}\leftarrow C_0$                                           \Comment{Current configuration}
		\State $C_{lastVoted}\leftarrow C_0$                                      \Comment{The last BMS configuration the node voted for}
		\State\label{ln:PendingRequests} $PendingRequests\leftarrow$ empty queue \Comment{Queue of pending valid join and leave requests}
		\State\label{ln:observedConfig} $ObservedConfig \leftarrow \{\}$         \Comment{Map from configurations to sets of nodes that observed them}
		\EndInit

		\State \textbf{upon received} $\langle\registerannounce,p\rangle$ \Comment{Check joining node registrations}
		\State \hskip 0.55cm \AndT $\exists r \in BMS.Registrations : r.id=p$ \textbf{ do}
		\State \hskip 0.55cm \hskip \algorithmicindent send $\langle\registerconfirm,p,\sigma_{id}\rangle$ to $p$
		\State

		\State\label{ln:RequestOrderingFirst} \textbf{upon receiving} $\langle\joinreq, p, pr\rangle$\textbf{ do} \Comment{Order join requests}
		\State \hskip 0.55cm $TotalOrderBroadcast.Broadcast(\langle\joinreq, p, pr\rangle)$
		\State

		\State \textbf{upon receiving } $\langle\leavereq,p,\sigma_p\rangle$\textbf{ do} \Comment{Order leave requests}
		\State \hskip 0.55cm $TotalOrderBroadcast.Broadcast(\langle\leavereq,p,\sigma_{id}\rangle)$
		\State

		\State \textbf{upon} $TotalOrderBroadcast.Deliver(\langle\joinreq, p ,pr\rangle)$                        \Comment{Check and enqueue}
		\State\label{ln:CheckProof} \hskip 0.55cm \textbf{ such that} $ValidProofOfRegistration(pr)$\textbf{ do} \Comment{ordered join requests}
		\State \hskip 0.55cm \hskip \algorithmicindent $PendingRequests.enqueue(\langle \joinreq, p \rangle)$
		\State

		\State \textbf{upon} $TotalOrderBroadcast.Deliver(\langle\leavereq, p ,\sigma_p\rangle)$     \Comment{Check and enqueue}
		\State\label{ln:CheckSig} \hskip 0.55cm \textbf{ such that} $ValidSig(\sigma_p)$\textbf{ do} \Comment{ordered leave requests}
		\State\label{ln:RequestOrderingLast} \hskip 0.55cm \hskip \algorithmicindent $PendingRequests.enqueue(\langle \leavereq, p \rangle)$
		\State

		\State\label{ln:BMSViewFirst}\textbf{upon} change of BMS$.C_{cur}$\textbf{ do}                                        \Comment{Order own observations}
		\State \hskip \algorithmicindent $TotalOrderBroadcast.Broadcast(\langle\textsc{BMS-CONFIG}$, BMS$.C_{cur},id\rangle)$ \Comment{of the BMS state}
		\State

		\UponCondition{TotalOrderBroadcast.Deliver(\langle \textsc{BMS-CONFIG},C,p\rangle)}      \Comment{Register other node's}
		\State\label{ln:BMSViewLast} $ObservedConfig[C] \leftarrow ObservedConfig[C] \cup \{p\}$ \Comment{observations of the BMS state}
		\EndUponCondition

		\UponCondition{length(PendingRequests) > 0 \wedge LastBMSConfig() \triangle C_{cur} < t}\label{ln:ReconfigFirst} \Comment{Reconfigure}
		\State $C_{cur}.number\leftarrow C_{cur}.number+1$
                \State $req \leftarrow PendingRequests.dequeue()$
		\If{$req = \langle \joinreq, p \rangle$}
		  \State $C_{cur}.members\leftarrow C_{cur}.members \cup \{p\}$
		  \State Respond to $p$
                \EndIf
		\If{$req = \langle \leavereq, p \rangle$}
		  \State $C_{cur}.members\leftarrow C_{cur}.members \setminus \{p\}$
                \EndIf\label{ln:ReconfigLast}
		\EndUponCondition

		\UponCondition{C_{lastVoted} \triangle C_{cur} \geq t}\label{ln:UpdateBMSFirst} \Comment{Vote for updating the configuration stored in BMS}
		\State send $\langle\vote,C,id\rangle$ to BMS
		\State $C_{lastVoted} = C_{cur}$
                \label{ln:UpdateBMSLast}
		\EndUponCondition

                \FunctionNoArgs{LatestBMSConfig}\label{ln:LatestBMSConfig}
		\State \textbf{return} $C$ with highest $C.number$ such that $|ObservedConfig[C] \cup C_{cur}| \geq f+1$
                \EndFunctionNoArgs

                \Function{ValidProofOfRegistration}{pr}
		\State \textbf{return} $|pr.confirmations| \geq f+1 \wedge \forall c \in pr.confirmations : ValidSig(c.\sigma_{id})$
                \EndFunction

	\end{algorithmic}
	\caption{Reconfiguration Protocol}
	\label{algorithm:ReconProt}
\end{algorithm*}

\subsubsection{Handling Reconfiguration Requests (Algorithm \ref{algorithm:ReconProt})}

The BFT system's replicated state contains, along with the application state, its current configuration $C_{cur}$.
Note that $C_{cur}$ might be ``ahead'' of the configuration stored by the reconfiguration service
(in case the configuration service has not yet been updated).
The BFT system also keeps track of all the nodes' local views of the configuration stored by BMS%
\footnote{If the reconfiguration service is implemented as an Ethereum smart contract,
  each node observes progressively newer states of the reconfiguration service
  by downloading more and more blocks of the Ethereum blockchain.
  This happens independently at each member of the BFT system.}
(line~\ref{ln:observedConfig}).
For this data to be consistent across all nodes,
they must agree on it before updating their copies of the BFT system's state (lines~\ref{ln:BMSViewFirst}-\ref{ln:BMSViewLast}).
We model the agreement by using a total order broadcast (TOB) abstraction, some form of which every BFT system possesses.
To account for $f(C_{cur})$ faulty nodes, we define the latest configuration stored in the BMS that the BFT system observed
as the one observed by at least $f(C_{cur}) + 1$ nodes of the current configuration (line~\ref{ln:LatestBMSConfig}).

Finally, the BFT system also stores a queue of pending reconfiguration requests ($\joinreq$ or $\leavereq$)
submitted by joining and leaving nodes (line~\ref{ln:PendingRequests}).
The nodes also keep this queue consistent by agreeing on its content using TOB (lines~\ref{ln:RequestOrderingFirst}-\ref{ln:RequestOrderingLast}).
They also discard invalid requests before inserting them in the queue (lines~\ref{ln:CheckProof} and \ref{ln:CheckSig}).
For brevity of the pseudocode, we assume that each request is delivered at most once by TOB,
even if broadcast by multiple nodes.

Once the BFT system reaches a state where
the queue of reconfiguration requests is not empty
and the configuration stored in BMS is not too different from the current configuration,
the BFT system processes the first pending reconfiguration request
and transitions to a new configuration that reflects the processed request (lines~\ref{ln:ReconfigFirst}-\ref{ln:ReconfigLast}).
We measure the difference between two configurations $C_j$ and $C_i$ as their symmetric difference $(C_i \triangle C_j)$,
which corresponds to the number of reconfiguration requests that need to be applied to $C_i$ in order to obtain $C_j$.
We consider two configurations too different if their difference reaches a threshold $t$, which is a system parameter.
In our evaluation we experiment with two different values for $t$:
$1$ and $f(C_{cur})/2$, where $f(C_{cur})$ is the tolerated number of faulty nodes in the current configuration\footnote{For simplicity of implementation, we use the current configuration and not the last published one for the computation of $t$. Using $t = C_{cur}/2$ is clearly below the bound on $t$ defined in Section~\ref{sec:batching}, regardless of the difference between $C_{cur}$ and the last published configuration.}.
To ensure that reconfiguration requests can be handled,
the BFT system updates the configuration stored in BMS
when the difference between its actual configuration and the configuration stored in BMS
reaches $t$ (lines~\ref{ln:UpdateBMSFirst}-\ref{ln:UpdateBMSLast}).

\section{Evaluation}
\label{sec:evaluation}

\subsubsection{Experimental Setup.} For each node, we use a virtual machine provided by a leading cloud provider,
with 4 CPUs and 8 GB memory, equipped with 1Gbps networking, running the Ubuntu 18.04 Linux distribution.
For the BFT system we use Mir-BFT \cite{stathakopoulou2019mir} configured as shown in Table~\ref{tab:mir}.
We use the hosted Ethereum node cluster Infura~\cite{infura} to run the reconfiguration service on Ropsten~\cite{ropsten},
an Ethereum test network. We evaluate primarily two aspects of our system: joining latency (Sec.~\ref{sec:evallatency}) and its cost (Sec.\ref{sec:evalcost}).

\begin{table}[h!]
  \centering
  \begin{tabular}{|r|l|}
    \hline
    Checkpoint Interval       & 20 s  \\ \hline
    Join Response Threshold   & $u_{cur} = f(C_{cur}) + 1$   \\ \hline
    Number of Peers & $4 \leq |C_{cur}| \leq 100$   \\ \hline
  \end{tabular}
  \caption{Mir-BFT configuration parameters used in the evaluation.
    In Mir-BFT, reconfiguration happens only when the system internally creates a checkpoint of its state.
     }
  \label{tab:mir}
  \vspace{-0.5cm}
\end{table}

\subsection{Joining Latency}
\label{sec:evallatency}

To evaluate the join latency for different system sizes,
we start with an initial configuration $C_0$ of Mir-BFT with $|C_0| = 4$ nodes.
We then make the system gradually grow by having new nodes join, one after the other,
until the system size reaches $|C_{96}| = 100$ nodes in its $96$-th configuration.
We configure the system to update the BMS state after each reconfiguration.
Only once the BMS state is updated, we add the next node. The adding of a node consists of 4 phases that contribute to the total join latency:
\begin{enumerate}
\item \label{enum:tx-latency} \textbf{Transaction latency:} The time between submission of the registration transaction by the joining node to BMS
  and the time this transaction becomes part of the BMS state.
  We consider the transaction to be part of the BMS state when it is included in a block of the Ropsten Ethereum testnet.
\item \label{enum:confirmation-latency} \textbf{Confirmation latency:} The time between the joining node requesting registration confirmation from the current members of the BFT system
  and the time when the joining node is able to send its join request to the BFT system.
  This phase involves the current members of the BFT system
  waiting for a sufficient number of blocks to consider the registration confirmed
  and sending a confirmation message to the joining node.
\item \textbf{Ordering latency:} Time it takes for Mir-BFT to order the reconfiguration request.
\item \textbf{Checkpoint latency:} Time between ordering the reconfiguration request and the next internal checkpoint
  at which Mir-BFT reconfigures and sends the final response to the joining node.
  We stop counting when the joining node receives the final response from $u_{cur} = f(C_{cur}) + 1$ nodes.
\end{enumerate}

\begin{figure*}[h]
  \centering
  \includegraphics[width=0.7\textwidth]{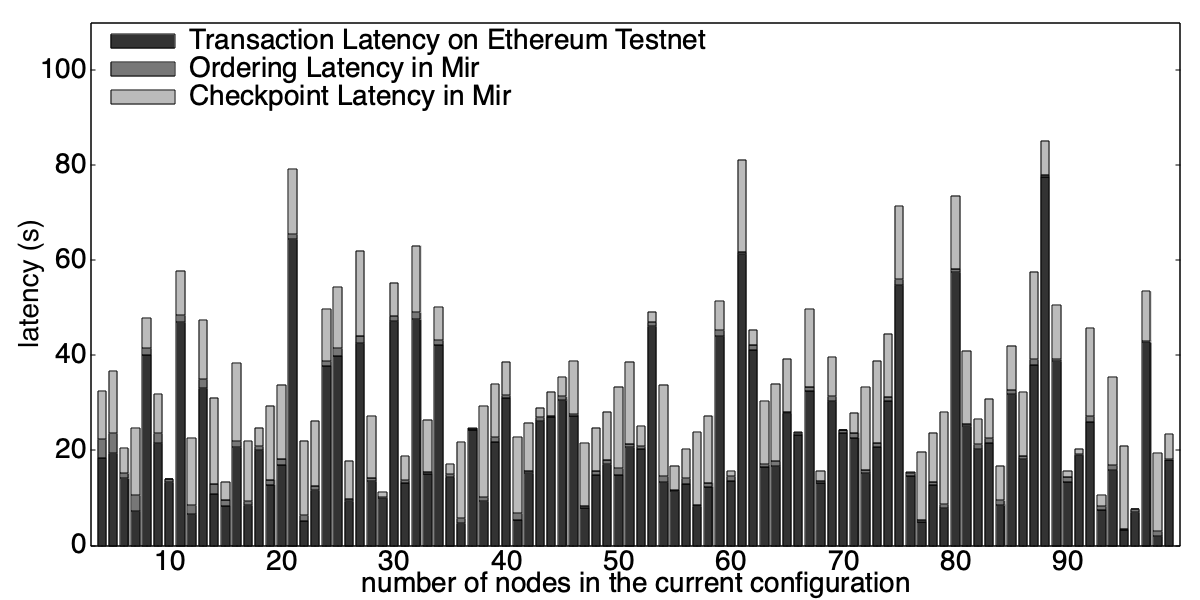}
  \caption{Breakdown of join latency for increasing system size (with $t=1$ policy).
  The dominating confirmation latency (of more than $9$ minutes) omitted for better readability.}
  \label{fig:join-t1-breakdown}
\end{figure*}

We measure the duration of these phases separately.
Figure \ref{fig:join-t1-breakdown} shows the results of one run from 4 to 100 nodes.
To speed up the experiment, we skip phase \ref{enum:confirmation-latency},
as it can be easily computed by multiplying the expected block creation time
by the number of confirmation blocks and adding one message round trip for the registration confirmation by the BFT system nodes.

\smallskip

\noindent\textbf{Transaction latency.}
The Ropsten Ethereum testnet has an average block generation time of $15 s$ with high variances.
Additionally, the transaction acceptance rate fluctuates significantly, e.g. due to network congestion~\cite{zhang2019ethereum}, as confirmed by
our measurements.
We observe an average transaction latency of $27.7 s$ with standard deviation of $24.9 s$.

\smallskip

\noindent\textbf{Confirmation latency.}
To match Bitcoin security, we assume that 37 Ethereum blocks are required to consider a transaction confirmed.
An expected block creation time of $15 s$ on the Ethereum network \cite{BlockTime}
thus results in a confirmation latency of $9.25 min$.
The round-trip time for the registration confirmation messages between the joining node and the members of the BFT system
is on the order of milliseconds and thus negligible.
The confirmation latency by far dominates the total joining latency
and we omit it from Figure \ref{fig:join-t1-breakdown} for better readability.

\smallskip

\noindent\textbf{Ordering and checkpoint latency.}
The latency of ordering a request in Mir-BFT is stable with increasing system size and,
with an average ordering latency of $0.95 s$, is not an influencing factor of the total join latency.
In our deployment, Mir-BFT reaches a checkpoint every $20 s$.
This is, as expected, the upper bound on the checkpoint latency that we observe.
The checkpoint latency stays between $0 s$ and $20 s$ as the request of a joining peer is,
in the best case, ordered right before the next checkpoint or,
in the worst case, right after a checkpoint.

\subsection{Joining Cost}
\label{sec:evalcost}

We now evaluate the cost associated with reconfiguration operations.
Given that we implement our reconfiguration service as an Ethereum smart contract,
the cost of executing associated transactions is measured in Gas.
We first evaluate the Gas cost of a single voting transaction and
the total amount of Gas consumed by a configuration update.
We then estimate as the cost for a single node to join the BFT system
both in terms of Gas and (using the current conversion rate) in US dollars.
This per-node join cost corresponds to the registration fee the joining node should pay.

We evaluate two policies for updating the configuration stored by the reconfiguration service.
The policy is expressed by the parameter $t$ -- the number of reconfigurations of the BFT system
after which it updates the reconfiguration service though the voting procedure.
The first policy updates BMS on every reconfiguration ($t=1$).
The second policy updates BMS after $f(C_{cur})/2$ reconfigurations ($t=f(C_{cur})/2$).
Note that $f(C_{cur})$, the number of tolerated failures, depends on the current configuration's size, and so does $t$ in this case.

\begin{figure*}[ht!]
  \begin{subfigure}{\columnwidth}
    \centering
    \includegraphics[width=\textwidth]{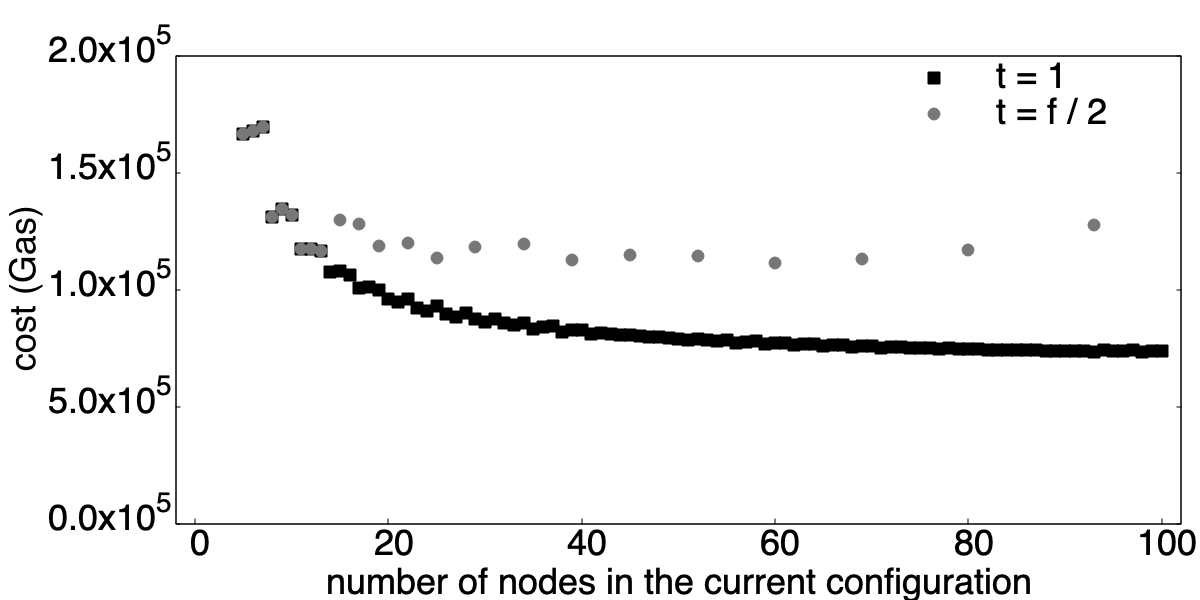}
    \caption{Average gas used by a single vote transaction.\\~\\~}
    \label{fig:avg-gas-per-vote}
  \end{subfigure}
\hspace{0.5cm}
  \begin{subfigure}{\columnwidth}
    \centering
    \includegraphics[width=\textwidth]{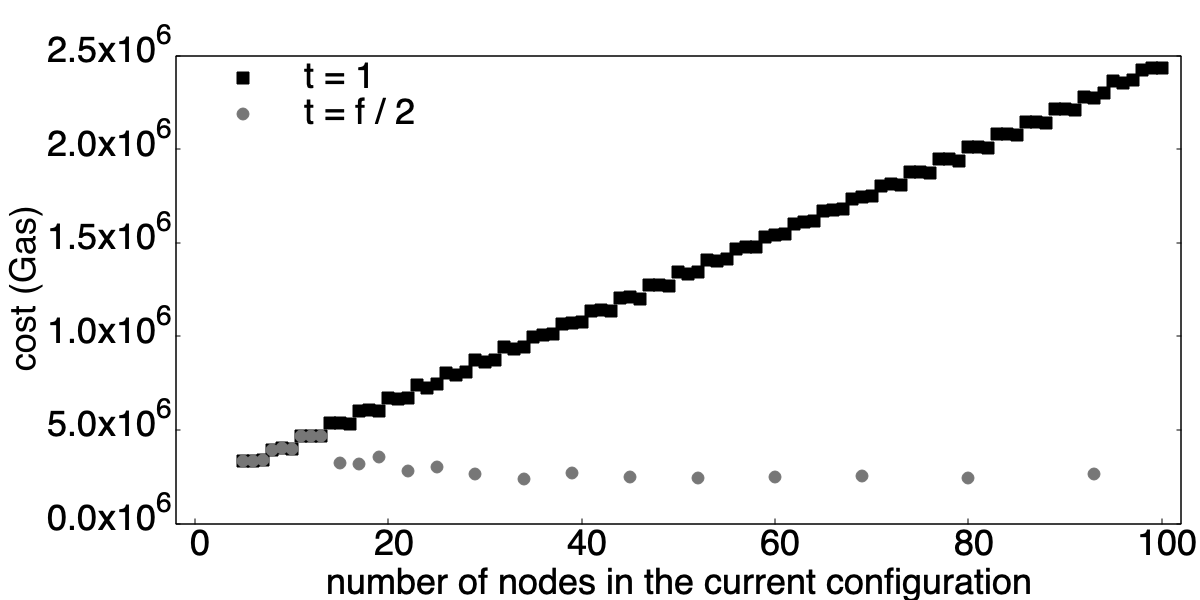}
    \caption{Total amount of gas used for a configuration update, divided by the joining peers included in the update for an increasing system size.}
    \label{fig:avg-gas-per-join}
  \end{subfigure}
\caption{Gas used by BMS with reconfiguration policies $t=1$ and $t=f(C_{cur})/2$}
\end{figure*}

\smallskip

\noindent\textbf{Cost of a Single Vote Transaction.}
Figure~\ref{fig:avg-gas-per-vote} compares the average amount of gas used by a single vote transaction for $t=1$ and $t=f(C_{cur})/2$.

For $t=1$, the average gas used for a single vote transaction decreases as the system size grows.
This is because the first vote transaction for a new configuration requires more Gas
for initializing certain data structures in the smart contract.
The last vote necessary to update the stored configuration is also more expensive
due to the extra computation induced by performing the configuration update.
In a bigger system, this increased cost is amortized over more simple vote transactions.

For $t=f(C_{cur})/2$, the transaction cost remains the same until the system is large enough for $t$ to increase from $1$ to $2$.
This occurs at $|C_{cur}|=13$, i.e., when $t = f(C_{cur}) / 2 = \lfloor((|C_{cur}| - 1) / 3)\rfloor / 2 = 2$.
The update itself being larger, it also requires more Gas to process than in the case of $t=1$.
Note the sparser data points for $t=f(C_{cur})/2$
caused by the voting occurring once every $t$ reconfigurations
rather than on every reconfiguration.

\smallskip

\noindent\textbf{Normalized Cost of a Configuration Update.}
Figure~\ref{fig:avg-gas-per-join} shows the the amount of gas consumed by a configuration update divided by the number of joining peers in that update.
We examine this value for an increasing system size and the reconfiguration policies $t=1$ and $t=f(C_{cur})/2$.

We observe that the total amount of gas increases linearly with the system size for the reconfiguration policy $t=1$.
This is due to the increasing number of votes $v=f(C_i)+1$ required to update the configuration, where $C_i$ the last published configuration. For $t=f(C_{cur})/2$, the amount of gas used for a configuration update
initially corresponds to the results for reconfiguration policy $t=1$
since, as discussed above, $t = f(C_{cur})/2 = 1$ in small systems.
For larger systems, however, the total cost of updating the configuration is divided by an increasing number of joining nodes
and the per-node joining cost remains low.

\smallskip

\noindent\textbf{Price of Joining in US Dollars.}
We now approximate the cost of adding a single node in US dollars based on the Gas measurements presented in Figure~\ref{fig:avg-gas-per-join}.

The Gas cost of a transaction is computed by multiplying its Gas price with the amount of used Gas.
As the Gas price is specified by the transaction sender,
we use the current average gas price of $93.1$ Gwei, determined using Etherscan (26.08.2020)~\cite{etherscan}.
To convert our results from Ether to US dollars, we use the current exchange rate of 1 Ether to $386.10$ US dollars (26.08.2020)~\cite{coinbase}.
It should be noted, however, that the exchange rate and average gas rate fluctuate heavily.
Table~\ref{tab:gasCost} shows the gas cost for a configuration update per joining node for different system sizes.
The registration fee for joining nodes should be calculated based on these values.
For a system size of $|C| \geq 19$, the current joining cost is approximately 4\$.

\begin{table}[h!]
		\centering
	\small{
	\begin{tabular}{|r|r|r|}
		\hline
		System Size & Amount of Gas & US Dollar \\ \hline
		5           & 166640        & 5.71      \\ \hline
		10          & 131970        & 4.52      \\ \hline
		15          & 129952        & 4.45      \\ \hline
		25          & 113314        & 3.88      \\ \hline
		45          & 114913        & 3.94      \\ \hline
		52          & 114460        & 3.92      \\ \hline
		60          & 111179        & 3.81      \\ \hline
		69          & 113146        & 3.88      \\ \hline
		80          & 117102        & 4.01      \\ \hline
		93          & 127590        & 4.37      \\ \hline
	\end{tabular}}
	\caption{Amount of gas and cost in USD for a configuration update per joining node}
	\label{tab:gasCost}
\end{table}

In addition to the cost of joining, each node's registration fee should also cover the cost of removal of that node from the configuration.
Removing nodes from a configuration being analogous to adding new ones, the cost of leaving is at most as high as the cost of joining.
We even expect leaving to be ``cheaper'' than joining, since Ethereum reduces the gas consumption of a transaction that frees storage space~\cite{wood2014ethereum}.

\section{BMS for Proof-of-Stake Systems and Extensions}
\label{sec:extensions}

\noindent\textbf{Reconfiguration of Proof of Stake Systems.} PoS blockchain systems inherently reconfigure their validator sets %
through the process of ``staking'' and ``destaking'' funds in native cryptocurrency.
To generalize BMS to PoS, we simply need to store in BMS,
in addition to the current configuration,
the amount of stake associated to each validator (i.e. member).
Each validator's vote for a new configuration then has a weight proportional to the size of the corresponding stake.

As discussed before,
destaking blocks the funds for weeks or months (``thawing time''),
in order to de-incentivize former validators from acting maliciously
towards clients that still believe them to be part of the validator set.
This is necessary because a client has no reliable means of distinguishing
between the new true validator set and a set of former validators that started acting maliciously.
With BMS, however, the client can make this distinction using the reconfiguration service.
Therefore, no incentive is needed to keep old validators well-behaved and they can reclaim their staked funds.
The advantage of using BMS instead of a long thawing time is thus twofold. First, the destaking time reduces from weeks or months (arbitrarily set threshold, during which all clients are assumed to update their view of the validator set) to the order of   minutes, as in our evaluation (Sec.~\ref{sec:evaluation}). Second,  long-range attacks, where a client reconnects ``too late'', are completely ruled out (unlike with---even very long---thawing times).

Another advantage of using BMS with a PoS blockchain is the possibility of rewarding nodes for interaction with BMS
in the native currency of the BFT system.
The interaction with BMS can be substantially simplified,
completely leaving out the registration of new nodes.
Instead, the amount of native currency a node has to lock as stake while joining
is higher than the amount returned to the node when leaving.
The difference corresponds to the registration fee.
A node that proves to have participated in the voting process can claim a part of this fee.
This approach does raise cryptocurrency exchange considerations, that are  out of the scope of this paper.

\smallskip

\noindent\textbf{Extensions: Hierarchical BMS and Multi-BMS.} Conceptually, BMS bears a similarity to Internet DNS (Domain Name System).
Apart from serving a similar purpose, BMS can even exhibit similar structural patterns.

Note that a \emph{fixed-membership BFT system} can be used instead of Ethereum in our system, as a root BMS.
Even PoS and dynamic-membership BFT systems can be used, as long as their configuration is registered in another BMS.
Such hierarchy can have arbitrary many levels and be rooted in a reliably discoverable system, that we call root BMS (i.e., a PoW blockchain or static-membership BFT system).

Moreover, analogously to the existence of multiple independent DNS services,
the configuration of a single BFT system
can rely on multiple independent BMS instances.
Upon reconfiguring, the BFT system would correspondingly update each of these BMSs.
A client would, depending on its trust in those reconfiguration services, use any / all / a quorum of them
to learn about the configuration of the BFT system.

\section{Related Work}
\label{sec:related}

\noindent\textbf{Reconfiguration of BFT Systems.}
Reiter~\cite{Reiter96membership}~\cite{Reiter96rampart} introduces a fundamental approach for reconfiguration by having the BFT system members agree on the next configuration with a 3-phase commit scheme.
Later efforts aim to improve the performance and scalability, e.g.
by optimizing fault-free execution~\cite{KihlstromMM01}, by having only a subset of nodes run the reconfiguration protocol~\cite{RodriguesLCLS12}, or by sharding membership updates across different regions~\cite{CowlingPLPG09}.
Similarly, CommChain~\cite{vizier2020comchain} blockchain addresses dynamic reconfiguration by having the set of nodes who run consensus, called deciders, periodically add a configuration block with the next set of deciders.
Such protocols, however, are vulnerable to ``I still work here'' attack \cite{aguilera2010reconfiguring}, that our BMS addresses, as joining and reconnecting nodes are not aware of the current set of nodes that implement the membership service.

Some BFT systems rely on centralization, e.g. require a system administrator to initiate the reconfiguration processes~\cite{BessaniSA14}.
Our solution replaces such centralized administrator with a decentralized BMS.

\smallskip

\noindent\textbf{Prevention of Long-range Attacks.}
A number of decentralized solutions have been proposed to prevent long-range and similar attacks in blockchain systems by introducing mechanisms to provide a finality layer.
Casper the Friendly Finality Gadget~\cite{buterin2017finalitycasper}, is a PoS based overlay system that enables the finalization of blocks in PoW Blockchains, e.g. Ethereum.
Casper
employs timestamps to enable protocol participants to ignore conflicting chains that revert finalized blocks.
This approach, however, addresses a slightly different (though related) problem of finality rather than preventing long-range attacks.
It still suffers from the issue that clients must know the current set of validators and thus go online frequently to track any changes.
Additionally, clients must verify the validator set at least once against a trusted source when they first connect to the system.

A recent theoretical result by Kuznetsov and Tonkikh~\cite{petrDisc2020}
prevents the ``I still work here'' / long range attack by using forward secure signatures,
without relying on a service external to the BFT system.
The nodes' responses to all requests, however, must be signed,
which may create a computational performance bottleneck in practical implementations.
Moreover, while the proposed solution only makes sure that an old configuration is unable to compromise a (re-)connecting client,
our approach can leverage the reconfiguration service to bootstrap clients' and joining nodes' initial views of the configuration.

Another recent proposal for PoS is  Winkle~\cite{AzouviDN19}, which leverages singed client transactions as weighted votes on recent blocks.
Blocks with a threshold of signatures constitute  checkpoints which cannot be equivocated even if their validators become corrupted.
However, checkpoints require clients being active, which, given current Bitcoin and Ethereum usage,
translates to delays in the order of months, even years.
Even using delegation, checkpoint delay remains in the order of hours or few days.
Moreover, Winkle relies on the assumption that an adversary cannot easily control a big number of clients.
Our solution, on the other hand, does not rely on validation from the clients and therefore clients can be arbitrarily corrupted.
Moreover, validators may join or leave the system within minutes, only constrained by the latency of updating the BMS state.

Tendermint/Cosmos~\cite{kwon2016cosmos}, a practical PoS system, currently employs a combination of measures to prevent long-range attacks.
Firstly, they make use of unbonding periods during which a deposits is made non-transferable to accounts.
Secondly, the first time a client connects to the network, it must verify the state of the chain against a trusted source,
and thirdly, clients are required to re-connect and track all changes in the validator set as at least as frequently as the unbonding period.
They have, however, also discussed other prevention mechanisms,
e.g., that non-validator token holders become trusted validity attestation sources for clients by posting their tokens as collateral with a much longer unbonding period.
Both options suffer from long periods during which coins are locked and require clients to frequently go online.
Our solution optimizes the thawing time and makes no assumptions on the up-time of clients.

\end{document}